\documentclass[wcp]{jmlr}
\usepackage{algorithmic}
\usepackage{algorithm}

\makeatletter
\let\Ginclude@graphics\@org@Ginclude@graphics 
\makeatother

\title[Neural Double Auction Mechanism]{Neural Double Auction Mechanism}

 \author{\Name{Tsuyoshi Suehara} 
 \addr Kyoto University
  \AND \Name{Koh Takeuchi} 
 \addr Kyoto University
  \AND \Name{Hisashi Kashima} 
 \addr Kyoto University
 \AND
 \Name{Satoshi Oyama} 
 \addr Nagoya City University
 \AND
 \Name{Yuko Sakurai} 
 \addr Nagoya Institute of Technology
 \AND
 \Name{Makoto Yokoo} 
 \addr Kyushu University
}

\editors{}

\begin{document}

\maketitle

\begin{abstract}
Mechanism design, a branch of economics, aims to design rules that can autonomously achieve desired outcomes in resource allocation and public decision making.
The research on mechanism design using machine learning is called automated mechanism design or mechanism learning.
In our research, we constructed a new network based on the existing method for single auctions and
aimed to automatically design a mechanism by applying it to double auctions.
In particular, we focused on the following four desirable properties for the mechanism: individual rationality, balanced budget, Pareto efficiency, and incentive compatibility.
We conducted experiments assuming a small-scale double auction and clarified how deterministic the trade matching of the obtained mechanism is. We also confirmed how much the learnt mechanism satisfies the four properties compared to two representative protocols.
As a result, we verified that the mechanism is more budget-balanced than the VCG protocol and more economically efficient than the MD protocol, with the incentive compatibility mostly guaranteed.
\end{abstract}

\section{Introduction}\label{sec-intro}

Mechanism design is a field of microeconomics that designs rules for the purpose of gaining better outcomes in resource allocation and public decision making, such as auctions, voting systems, and matching problems.
Various studies have long been conducted; for example,   \cite{art7} realized an optimal auction design for a single item setting.
There have been other approaches for automated mechanism design, which is achieved by solving the optimization problem with computational method~\citep{proc0,proc11,art8}.
They are however limited to specific auction classes known to be incentive compatible. 

As a more recent research for designing optimal auctions, \cite{proc1} proposed a novel mechanism learning framework by using deep learning architectures. They developed \textit{RegretNet}, a feed-forward neural network, to approximate optimal results in known multi-item settings and found new mechanisms in unknown cases. 
Despite the rapid development of mechanism learning, there still lacks an automated mechanism design through deep learning for essential types of auction, such as double auction, which considers more complex settings than a common single auction.

\begin{figure}[t]
 \centering
  \begin{minipage}[b]{0.45\linewidth}
    \centering
    \includegraphics[keepaspectratio, scale=0.65]{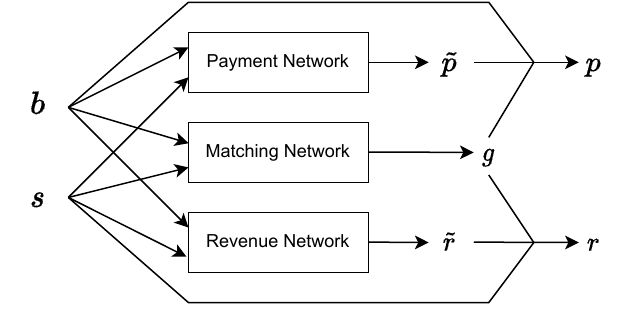}
    {(a) \textit{DoubleRegretNet} (proposed)}
  \end{minipage}
  \quad
  \begin{minipage}[b]{0.45\linewidth}
    \centering
    \raisebox{8mm}{\includegraphics[scale=0.65]{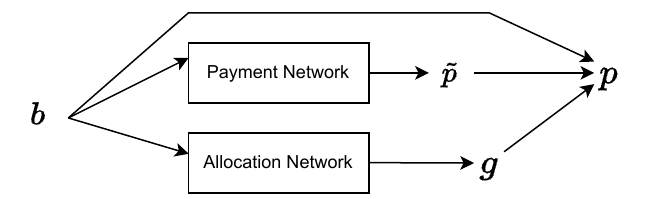}}
    {(b) \textit{RegretNet}}
  \end{minipage}
  \caption{The model architecture of the proposed (a) \textit{DoubleRegretNet} compared with the existing (b) \textit{RegretNet}.
  \textit{DoubleRegretNet} consists of three components: the payment, matching, and revenue networks to deal with a double auction.
  $b$ and $s$ are the valuations of buyers and sellers, respectively. Buyer-seller matching, payments, and revenues are denoted by $g$, $p$, and $r$, respectively.}
  \label{fig}
\end{figure}

Double auction, in which both the sellers and buyers report their valuations, is suitable for trading a large number of the same commodity and is used in securities trading, foreign exchange trading, and cryptocurrency trading, such as Bitcoin.
Double auction can also be applied to the supply chain, where each single market forms a link between buyers in the market and sellers in the next market~\citep{proc12}.
Previous studies proposed some protocols for double auction from various viewpoints of desirable properties, but there have been no automatically designed mechanisms~\citep{art5,proc6,proc12}.

In this study, we aimed to realize an automated mechanism design for double auction through deep learning.
We first formulated a double auction as an optimization problem that incorporates balanced budget and incentive compatibility into constraints with the overall utility maximized as much as possible.
To solve the optimization problem, we propose a novel symmetric learning network called \textit{DoubleRegretNet} based on \textit{RegretNet} to determine who sells or buys an item at what price when bid profiles of players are given as inputs.
We show an overview of the existing and proposed network architectures for mechanism learning in Figure \ref{fig}. 
Different from \textit{RegretNet}, our proposed method consists of the payment, matching, and revenue networks, which will be introduced in the following section.
We devise a structure to learn a mechanism that is equivariant to the permutation of inputs and guarantees individual rationality.

To compare our proposed method with existing MD and VCG protocols, we conducted experiments in a small-scale double auction. 
The results show how deterministic the learnt trade matching is and how much the mechanism meets four desirable properties.
We show the visualization of outputs of three networks in \textit{DoubleRegretNet} and demonstrate that the learnt mechanism indicates new insights about the double auction mechanism.
Moreover, as a result of the numerical experiment, \textit{DoubleRegretNet} learns a specific mechanism that brings more social surplus than the MD protocol and reduces the auctioneer deficit more than the VCG protocol.

We summarize the main contribution of this study as follows:
\begin{itemize}
    \item 
    We formulate a single unit double auction as an optimization problem considering balanced budget, Pareto efficiency, and incentive compatibility.  
    \item 
    We propose a symmetric learning network called \textit{DoubleRegretNet} to learn an individually rational mechanism for double auction.
    \item 
    We demonstrate the potential of automatically designing a new double auction mechanism by visualizing the outputs of neural networks in a particular setting.
    \item 
    We evaluate the validity of the learnt mechanism in several settings to compare with existing protocols. We actually obtain a mechanism that brings high utilities while almost satisfying balanced budget and incentive compatibility.
    
\end{itemize}

\section{Related Work}
By following the seminal work by \cite{proc1} which proposed \textit{RegretNet} to discover optimal auctions using deep learning,
several researchers have attempted to improve or extend \textit{RegretNet} for better performance~\citep{proc9} and for dealing with various settings in mechanism design, such as \textit{EquivariantNet} considering the symmetry of buyers~\citep{proc2} and \textit{Deep Neural Auction} for large scale end-to-end learning of e-commerce advertising auctions~\citep{proc3}.
Extensions to other settings include false-name-proof auctions~\citep{proc4}, budget constraints of players~\citep{proc5}, and matching problems~\citep{art1}, to name a few.
These studies leverage the expressive power of deep neural networks to approach more complex problems.
More recently, \textit{Regret-Former}~\citep{art9} took a similar approach to RegretNet, attempting to minimize dominant-strategy incentive compatibility violation.
\cite{art10} proposed \textit{AMenuNet} that strictly guarantee strategy-proofness by focusing on affine-maximizing Vickrey-Clarke-Groves mechanisms.

\section{Background}

\subsection{Preliminaries} 

We consider a single-unit double auction with a set of $n$ buyers $N = \{1, ..., n\}$, set of $m$ sellers $M = \{1, ..., m\}$, and auctioneer who is neutral to both buyers and sellers. Each buyer $i \in N$ has a truthful valuation $v_i^{\text{(B)}} \in \mathbb{R}_{\ge{0}}$, and each seller $j \in M$ has a truthful valuation $v_j^{\text{(S)}} \in \mathbb{R}_{\ge{0}}$. 
The valuation of the $i$-th buyer and $j$-th seller are independently drawn from distributions $F_i^{\text{(B)}}$ and $F_j^{\text{(S)}}$, respectively.

The auctioneer knows the valuation distributions $F^{\text{(B)}} = (F_1^{\text{(B)}}, ..., F_n^{\text{(B)}} )$ and $F^{\text{(S)}} = (F_1^{\text{(S)}}, ..., F_m^{\text{(S)}} )$ but does not know the realized valuations of buyers and sellers, which are denoted as $v^{(B)} = (v_1^{\text{(B)}}, ..., v_n^{\text{(B)}} )$ and $v^{(S)} = (v_1^{\text{(S)}}, ..., v_m^{\text{(S)}} )$, respectively.
In a double auction, the buyers and sellers report their valuations $b = (b_1, ..., b_n)$ and $s = (s_1, ..., s_m)$ , which are perhaps untruthful to the auctioneer.
Afterwards, the auction decides which player should trade and at what price.

In general double auctions, whether or not a buyer $i$ and seller $j$ make a deal is represented by a binary value $\{0$:no$, 1$:yes$\}$.
However, because binary values are difficult to learn by neural network models, we instead introduce randomized trade matching $g_{ij} \in [0, 1]$ by following a common assumption that each player makes a deal at most once. Let us denote a set of randomized trade matching as an $m \times n$-matrix $g$ that satisfies the following conditions:

\begin{align}
\begin{split}
\label{cond_g}
     g_{ij} \in [0, 1]\quad  \text{s.t.}\quad &\sum_{k=1}^m g_{ik} \le 1,  \forall i \in N,  \quad \sum_{k=1}^n g_{kj} \le 1, \forall j \in M.
\end{split}
\end{align} 
We denote the buyer payments and seller revenues by $p = (p_1, p_2, ..., p_n)$ and $r = (r_1, r_2, ..., r_m)$, which are $n$ and $m$-dimensional non-negative real vectors, respectively. 
Based on the above definitions, we define a result of a double auction as $(g, p, r)$, which is a tuple of trade matching, payments, and revenues.

After the result $(g, p, r)$ is decided, a buyer with the valuation $v_i^{\text{(B)}}$ receives the utility $u_i^{\text{(B)}}(v_i^{\text{(B)}}; b, s) = v_i^{\text{(B)}}\sum_{j=1}^m g_{ij}(b, s) - p_i(b, s)$ at buyer bid profile $b$ and seller bid profile $s$. 
In the same way, a seller with valuation $v_j^{\text{(S)}}$ receives the utility $u_j^{\text{(S)}}(v_j^{\text{(S)}}; b, s) = r_j(b, s) - v_j^{\text{(S)}}\sum_{i=1}^n g_{ij}(b, s)$. 
We define the utility function of auctioneer $u^{\text{(A)}}$ as the difference between the sum of the buyer payment and that of the seller revenue, $u^{\text{(A)}} = \sum_{i=1}^n p_{i} - \sum_{j=1}^m r_{j}$.

Here, we introduce four properties of double auction that a mechanism should satisfy to avoid unreasonable losses for participants.

Let us denote the valuation profile without the $i$-th element $v_i^{\text{(B)}}$ as $v_{-i}^{\text{(B)}}$, and similarly for $v_{-j}^{\text{(S)}}, b_{-i},$ and $s_{-j}$.
We consider a double auction as \textit{individually rational} (IR) if each player receives a non-negative utility when
participating in the auction truthfully. Let us define the IR property as

\begin{align*} 
  u_i^{\text{(B)}}(v_i^{\text{(B)}}; (v_i^{\text{(B)}}, b_{-i}), s) &= v_i^{\text{(B)}}\sum_{j=1}^m g_{ij} - p_i \ge 0, 
 \quad \forall i \in N,\\ 
  u_j^{\text{(S)}}(v_j^{\text{(S)}}; b, (v_j^{\text{(S)}}, s_{-i})) &= r_j - v_j^{\text{(S)}}\sum_{i=1}^n g_{ij} \ge 0,\quad \forall j \in M.
\end{align*}
In addition, an auction is (weakly) budget-balanced (BB) if the auctioneer receives a non-negative utility

\begin{align*} 
  u^{\text{(A)}} = \sum_{i=1}^n p_{i} - \sum_{j=1}^m r_{j} \ge 0.
\end{align*}
BB indicates that the auctioneer does not receive a loss by holding an auction.
We regard an auction as Pareto efficient (PE) if no trade matching is available that makes one individual better off without making another worse off. In our problem setting, Pareto efficiency is equivalent to the maximization of social surplus \textbf{welfare}, and we denote the sum of utilities in a double auction by
\begin{eqnarray}
\label{eq:welfare}
  \textbf{welfare} &=&\sum_{i=1}^n u_i^{\text{(B)}} + \sum_{j=1}^m u_j^{\text{(S)}} + u^{\text{(A)}} \nonumber\\
          &=&\sum_{i=1}^n \Bigl(v_i^{\text{(B)}}\sum_{j=1}^m g_{ij} - p_i\Bigr) + \sum_{j=1}^m\Bigl(r_j - v_j^{\text{(S)}}\sum_{i=1}^n g_{ij}\Bigr) + \Bigl(\sum_{i=1}^n p_{i} - \sum_{j=1}^m r_{j}\Bigr)\nonumber\\
          &=&\sum_{i=1}^n \Bigl(v_i^{\text{(B)}}\sum_{j=1}^m g_{ij}\Bigr) - \sum_{j=1}^m\Bigl(v_j^{\text{(S)}}\sum_{i=1}^n g_{ij}\Bigr).
\end{eqnarray}
Finally, we regard an auction as dominant strategy incentive compatible (DSIC) if each player utility is maximized by reporting
truthfully no matter what the other players report:

\begin{align*}
  u_i^{\text{(B)}}(v_i^{\text{(B)}}; (v_i^{\text{(B)}}, b_{-i}), s) &\ge u_i^{\text{(B)}}(v_i^{\text{(B)}}; (b_i, b_{-i}), s), \quad \forall i \in N, \\ 
  u_j^{\text{(S)}}(v_j^{\text{(S)}}; b, (v_j^{\text{(S)}}, s_{-i})) &\ge u_j^{\text{(S)}}(v_j^{\text{(S)}}; b, (s_j, s_{-i})), \quad \forall j \in M.
\end{align*}
This condition means that, in a DSIC auction, the optimal strategy for each player is to honestly report its truthful valuation.
In other words, we can ensure that reporting untruthful valuations offers no benefits to buyers and sellers.

\subsection{Existing Protocols}
~\cite{art6} proved that there is no mechanism that simultaneously satisfies all of the properties, which are introduced in the previous section, including the individual rationality, balanced budget, incentive compatibility, and Pareto efficiency for trading a single good in a double auction. 
Therefore, we present an existing protocol that guarantees three of the four properties.

The Vickrey-Clarke-Groves protocol (VCG
protocol)~\citep{art2,art3,art4} is a protocol where a player pays the reduced utility by their participation in an auction. When assuming efficient allocation, this is also thought of as paying the sum of damaged utilities of other players by that player joining the auction. This mechanism guarantees three properties other than balanced budget.

McAfee analyzed dominant strategies for buyers and sellers in double auction~\citep{art5}.
McAfee's Double Auction protocol (MD protocol) is a protocol that
guarantees three properties except Pareto efficiency.

\section{Proposed Method}
\subsection{Double Auction Mechanism Design as a Learning Problem}
As stated in the previous section, the auctioneer does not know truthful valuations of buyers and sellers. 
Therefore, we need to develop a mechanism that determines outputs only from the observed bid profiles $b$ and $s$. 
In the mechanism design, there are various targets for a mechanism, such as maximizing the profit of an auctioneer, buyers, or sellers.
In this study, to achieve an auction result that is closer to Pareto efficient for all participants but not for only a subset of participants, we define an objective function to maximize the sum of utilities that is defined as \textbf{welfare} in Equation \eqref{eq:welfare}.

Then, we introduce constraints to guarantee the incentive compatibility as much as possible. 
We define regrets of buyers and sellers as measures of the extent to which the mechanism violates incentive compatibility. 
The regret is the maximum increase in each player utility by considering arbitrary lies,

\begin{align*}
&\textbf{rgt}_i^{\text{(B)}}(w) = \mathbb{E}[\max_{\quad {v'_i}^{\text{(B)}}} u_i^{\text{(B)}}(v_i^{\text{(B)}}, ({v'_i}^{\text{(B)}}, {v^{\text{(B)}}_{-i}}), v^{\text{(S)}}) - u_i^{\text{(B)}}(v_i^{\text{(B)}}, v^{\text{(B)}}, v^{\text{(S)}})],\\ 
&\textbf{rgt}_j^{\text{(S)}}(w) = \mathbb{E}[\max_{\quad {v'_j}^{\text{(S)}}} u_j^{\text{(S)}}(v_j^{\text{(S)}}, v^{\text{(B)}}, ({v_j}^{\text{(S)}}, {v^{\text{(S)}}_{-j}})) - u_j^{\text{(S)}}(v_j^{\text{(S)}}, v^{\text{(B)}}, v^{\text{(S)}})].
\end{align*}
These express the expected values for $v^{\text{(B)}} \sim F^{\text{(B)}}$ and $v^{\text{(S)}} \sim F^{\text{(S)}}$. The mechanism satisfies the incentive compatibility only if $\textbf{rgt}_i^{\text{(B)}}(w)=0, \forall i \in N$ and $\textbf{rgt}_j^{\text{(S)}}=0, \forall j \in M$.

We also should care about the balanced budget in an auction. It is naturally desirable for the auctioneer to have a mechanism to avoid losing as much money as possible. We denote \textbf{bbp} as the penalty for balanced budget, i.e.,

\begin{align*}
\textbf{bbp}(w) = \mathbb{E} \biggl[\max \Bigl\lbrace 0, \sum_{j=1}^m r_{j}^w(v^{\text{(B)}}, v^{\text{(S)}}) - \sum_{i=1}^n p_{i}^w(v^{\text{(B)}}, v^{\text{(S)}}) \Bigr\rbrace \biggr].
\end{align*}
This shows that the auctioneer has to bear the difference between the sum of buyer payments and sum of sellers revenues. The mechanism satisfies balanced budget only if $\textbf{bbp}(w)=0$.

In our study, we realize the individually rational mechanism by incorporating constraints on payments and revenues into a neural network architecture. The details are discussed in the next section.

We formulate the double auction design as a learning problem based on the above discussion.
Given $L$ valuation profiles from $F^{(B)}$ and $F^{(S)}$ for mini-batch learning, we need to compute player regrets and budget-balanced constraint violations. Thus, we rewrite the penalty as follows:

\begin{align*}
\widehat{\textbf{rgt}}_i^{\text{(B)}}(w) &= \frac{1}{L} \sum_{\ell=1}^L \max_{\quad {v^{\prime\text{(B)}}_{\ell, i}}} u_{i}^{\text{(B)}}(v_{\ell, i}^{\text{(B)}}, ({v^{\prime\text{(B)}}_{\ell, i}}, {v^{\text{(B)}}_{\ell, -i}}), v_{\ell}^{\text{(S)}}) - u_{i}^{\text{(B)}}(v_{\ell, i}^{\text{(B)}}, v_{\ell}^{\text{(B)}}, v_{\ell}^{\text{(S)}})\\ 
\widehat{\textbf{rgt}}_j^{\text{(S)}}(w) &= \frac{1}{L} \sum_{\ell=1}^L\max_{\quad {v^{\prime\text{(S)}}_{\ell, j}}} u_{j}^{\text{(S)}}(v_{\ell, j}^{\text{(S)}}, v_{\ell}^{\text{(B)}}, ({v^{\prime\text{(S)}}_{\ell, j}}, {v^{\text{(S)}}_{\ell, -j}})) - u_{j}^{\text{(S)}}(v_{\ell, j}^{\text{(S)}}, v_{\ell}^{\text{(B)}}, v_{\ell}^{\text{(S)}})\\ 
\widehat{\textbf{bbp}}(w) &= \frac{1}{L} \sum_{\ell=1}^L \max \Bigl\lbrace 0, \sum_{j=1}^m r_{j}^w(v_\ell^{\text{(B)}}, v_\ell^{\text{(S)}}) - \sum_{i=1}^n p_{i}^w(v_\ell^{\text{(B)}}, v_\ell^{\text{(S)}}) \Bigr\rbrace.
\end{align*}
As a result, the optimization problem we try to solve is
\begin{equation}
\begin{split}
  \label{opt}
  \displaystyle \min_{w \in \mathbb{R}^d} \quad -&\frac{1}{L} \sum_{\ell=1}^L\Bigl( \sum_i^n (v_{\ell, i}^{\text{(B)}}\sum_j^m g_{ij}^w) - \sum_j^m(v_{\ell, j}^{\text{(S)}}\sum_i^n g_{ij}^w) \Bigr) \\
  \text{s.t.} \qquad &\widehat{\textbf{rgt}}_i^{\text{(B)}}(w) = 0, \quad \forall i \in N, \\
              &\widehat{\textbf{rgt}}_j^{\text{(S)}}(w) = 0, \quad \forall j \in M, \\
              &\widehat{\textbf{bbp}}(w) = 0.
\end{split}
\end{equation}

\subsection{Model Architecture}
\begin{figure}[t]
  \includegraphics[scale= 0.7]{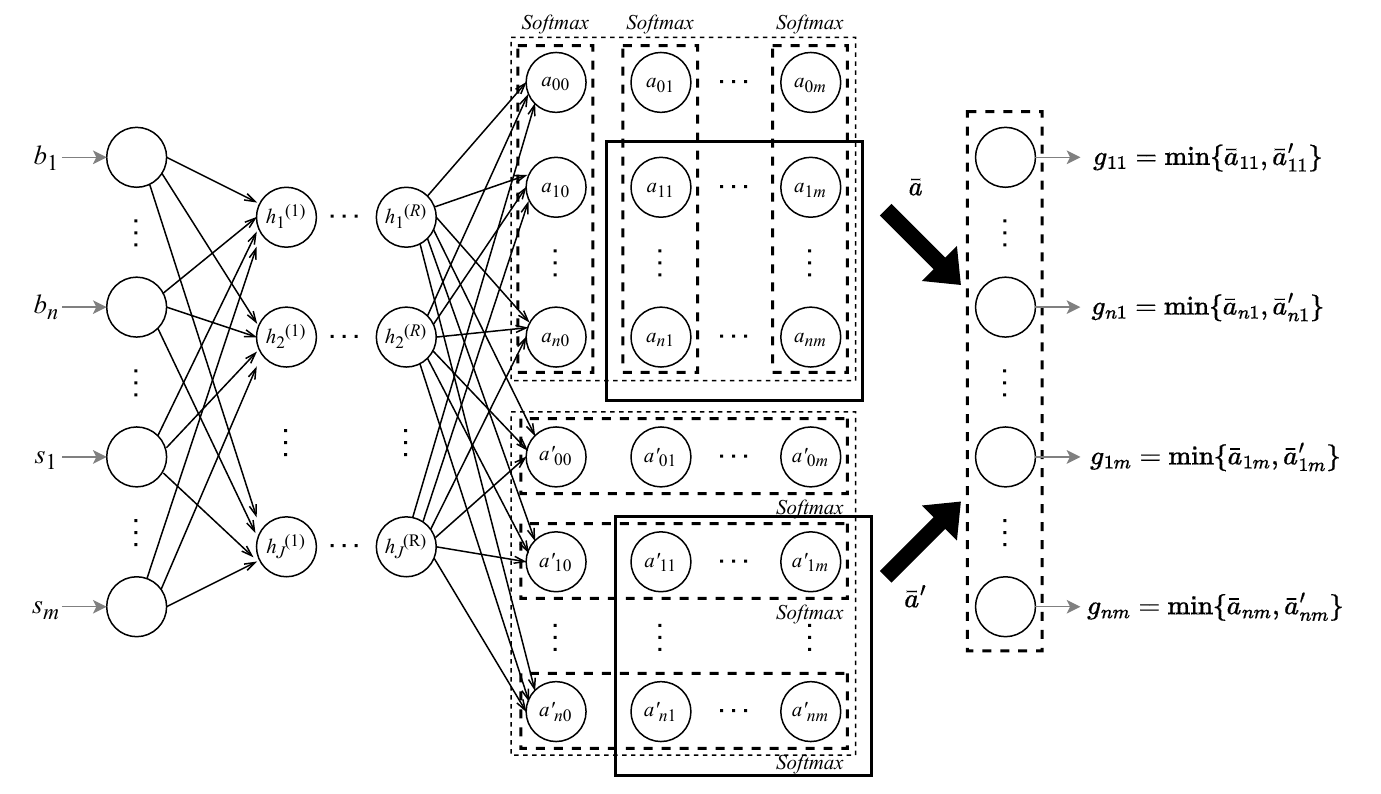}
  \caption{Matching Network}
  \label{fig1}
\end{figure}

Here, we describe \textit{DoubleRegretNet}, a neural network architecture for learning double auction mechanisms. 
As shown in Figure \ref{fig}, this architecture consists of three different elements: a matching network, payment network, and revenue network.

A matching network is a network for learning of whether one buyer and one seller will trade. 
We replace the inputs of the allocation component in \textit{RegretNet}~\citep{proc1} and call it a matching network in this paper.
Figure \ref{fig1} shows an overview of the matching network architecture. The network inputs the bids of players and outputs the probability of making a deal $g_{ij} \in [0, 1]$ for any buyer $i \in N$ and seller $j \in M$. After taking the input, it computes the $(n+1) \times (m+1)$-dimensional matrices $a$ and $a'$ via several fully connected layers.
Then, we apply the softmax function to two computed matrices; one is calculated along each row, and the other is calculated along each column.
Through this, we ensure the condition that each player trades at most one good.
The dimensions of the rows and columns are increased by one to account for the case where the player does not trade with anyone.
In Figure \ref{fig1}, $a_{i0}$ ($a_{0j}$) as dummy nodes hold the probability that the buyer $i$ (seller $j$) does not make a deal.
We gain the output $n \times m$-dimensional matrix $g$ obtained by smaller element of $\overline{a}$ and $\overline{a}'$, i.e.,

\begin{align*}
g_{ij} = \min \Bigg\{\dfrac{e^{a^{}_{ij}}}{\sum_{k=0}^n e^{a_{kj}}}, \dfrac{e^{a'_{ij}}}{\sum_{k=0}^m e^{a'_{ik}}} \Bigg\}.
\end{align*}
This is why the trade matching $g$ meets the condition (\ref{cond_g}).

\begin{figure}[t]
  \includegraphics[scale= 0.7]{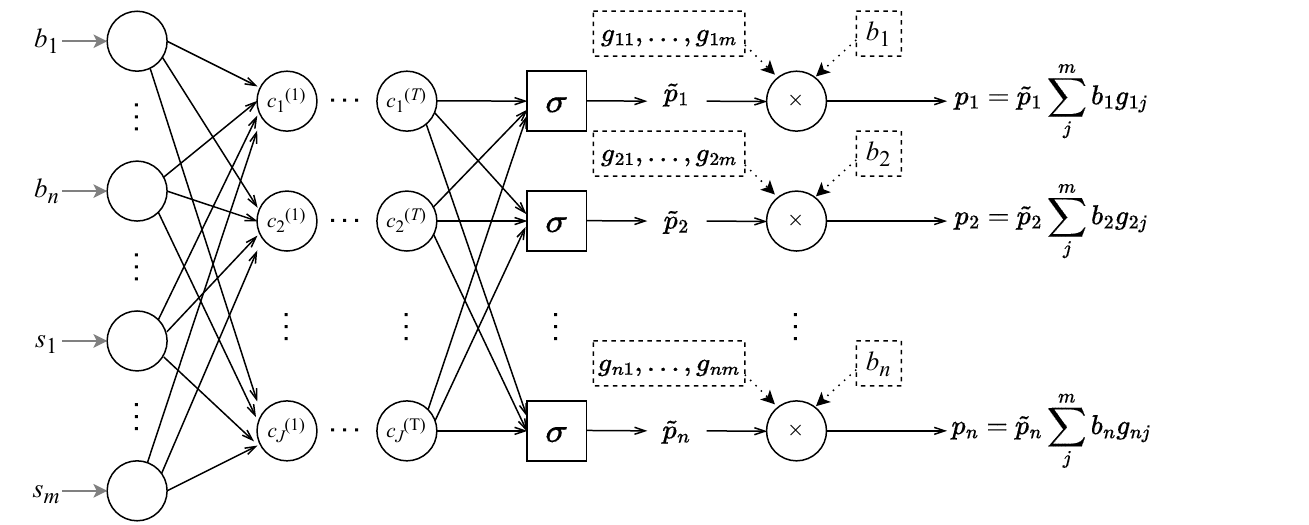}
  \caption{Payment Network}
  \label{fig2}
\end{figure}

The payment network~\citep{proc1} is a network for learning how much a buyer pays. 
Figure \ref{fig2} provides an overview of the payment network architecture, with the inputs changed for double auction. The network inputs the bids of players and outputs the amount $p_i$ paid by buyer $i$. Like the matching network, the payment network calculate specific values through fully connected layers. 
We make the learnt mechanism ensure the buyer individual rationality by applying the sigmoid function to the obtained values and computing the normalized payment $\tilde{p}_i \in [0, 1]$.
Then, the network decides the actual payments by calculating $p_i = \tilde{p}_i \sum_{j=1}^m b_i g_{ij}$.
The $g_{ij}$ in this formula is obtained from the matching network.
As a result, the amount each buyer has to pay is definitely less than or equal to their reported bid. 

\begin{figure}[t]
  \includegraphics[scale= 0.7]{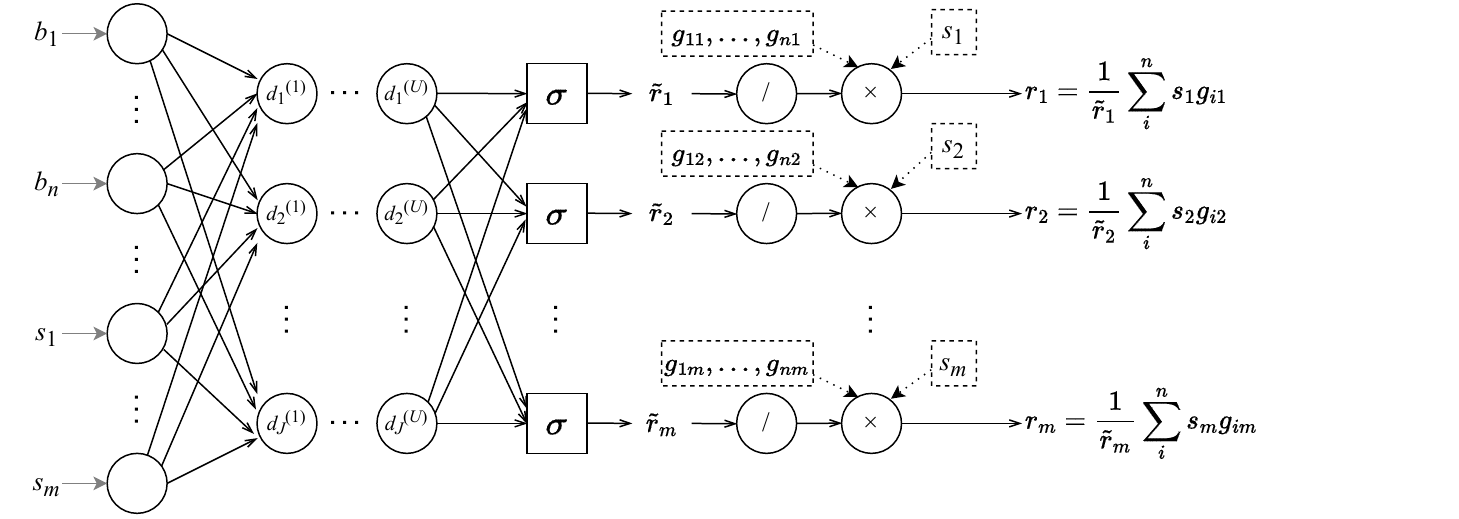}
  \caption{Revenue Network}
  \label{fig3}
\end{figure}

We construct the revenue network, similar to the payment network, for learning how much a seller receives. 
Figure \ref{fig3} presents an overview of the revenue network architecture, similar to the payment network. 
To ensure that the auction satisfies the seller individual rationality, the network first computes a normalized revenue $\tilde{r}_j \in [0, 1]$ for each seller $j$ using the sigmoid function.
Then, the network outputs an actual revenue $r_j = \frac{1}{\tilde{r}_j} \sum_{i=1}^n s_j g_{ij}$, where we can obtain $g_{ij}$ from the trade matching component.
This guarantees individually rational mechanisms for sellers because the amount each seller should receive is absolutely more than or equal to their reported bid. 

Moreover, we consider the symmetry of this problem setting for a double auction. We do not deal with any information or attributes of the players other than the bid profiles.
We also assume that the distributions $F_i^{\text{(B)}}$ and $F_{i'}^{\text{(B)}}$ ($F_j^{\text{(S)}}$ and $F_{j'}^{\text{(S)}}$) that the valuations of the buyer $i$ and $i'$ (seller $j$ and $j'$) follow are identical for any buyer $i, i' \in N$ (for any seller $j, j' \in M$). 
Therefore, it is natural that the auction result should be independent of the input order if the player bids are the same as a set.
In other words, it is desirable to learn a mechanism that is equivariant to the permutation of the inputs considering the symmetry among players.
We apply operations to the inputs of the network to gain an equivariant double auction mechanism. Specifically, we preliminarily sort the sampled values $b = (b_1, ..., b_n)$ and $s = (s_1,..., s_m)$ in descending order and give them to each network as inputs.
We obtain the output results from sorted inputs and finally rearrange them to correspond to the indices before sorting.

\subsection{Training}

We next describe how to train the neural network architectures.
We use the augmented Lagrangian method to solve the constrained optimization problem in (\ref{opt}) over the space of neural network parameters $w$. The Lagrangian function $C_p(w;\lambda)$, which can be expressed as the objective function plus a quadratic penalty term for constaraint violations, is

\begin{align*}
    C_p(w;\lambda) =& -\frac{1}{L} \sum_{\ell=1}^L\Bigl( \sum_i^n (v_{\ell, i}^{\text{(B)}}\sum_j^m g_{ij}^w) - \sum_j^m(v_{\ell, j}^{\text{(S)}}\sum_i^n g_{ij}^w) \Bigr)\\
                   &+ \sum_{i\in N} \lambda_i^{\text{(B)}} \widehat{\textbf{rgt}}_i^{\text{(B)}}(w) + \frac{\rho}{2}\sum_{i \in N} \Bigl(\widehat{\textbf{rgt}}_i^{\text{(B)}}(w)\Bigr)^2 \\
                   &+ \sum_{j\in M} \lambda_j^{\text{(S)}} \widehat{\textbf{rgt}}_j^{\text{(S)}}(w) + \frac{\rho}{2}\sum_{j \in M} \Bigl(\widehat{\textbf{rgt}}_j^{\text{(S)}}(w)\Bigr)^2 \\
                   &+ \lambda^{\text{(A)}} \widehat{\textbf{bbp}}(w) + \frac{\rho}{2} \Bigl(\widehat{\textbf{bbp}}\,(w)\Bigr)^2,
\end{align*} 
where $\lambda^{\text{(B)}} \in \mathbb{R}^n, \lambda^{\text{(S)}} \in \mathbb{R}^m$, and $\lambda^{\text{(A)}} \in \mathbb{R}$ are Lagrange multipliers, and $\rho > 0$ is a fixed parameter that controls the weight on the quadratic penalty.
We alternately update the model parameters and Lagrange multipliers as follows:

\begin{align*}
    &(a)\quad w^{\text{new}} \in \text{argmin}_w\; \mathcal{C}_p(w^{\text{old}}; \lambda^{old})\\
    &(b)\quad {\lambda_i^{\text{(B)}}}^{\text{new}} = {\lambda_i^{\text{(B)}}}^{\text{old}} + \rho \; \widehat{\textbf{rgt}}_i^{\text{(B)}}(w^{\text{new}}), \forall i\in N \\
    & \qquad \,  {\lambda_j^{\text{(S)}}}^{new} \hspace{0.1mm}= {\lambda_j^{\text{(S)}}}^{\text{old}} \hspace{0.5mm}+ \rho \; \widehat{\textbf{rgt}}_j^{\text{(S)}}(w^{\text{new}}), \forall j \in M\\
    & \qquad \,  {\lambda^{\text{(A)}}}^{\text{new}} = {\lambda^{\text{(A)}}}^{\text{old}} + \rho \; \widehat{\textbf{bbp}}\,(w^{\text{new}}).
\end{align*}

\textbf{Algorithm \ref{alg1}} describes the detailed learning algorithm.
The training sample $\mathcal{S}$ is divided into mini-batches of size L, and these are used as
training data multiple times (the data are randomly shuffled after each iteration). Then, we denote the mini-batches obtained in the $t$-th iteration as $\mathcal{S}_t = \{(v_1^{\text{(B)}}, v_1^{\text{(S)}}), ..., (v_L^{\text{(B)}}, v_L^{\text{(S)}})\}$.
The update $(a)$ of the parameter $w$ corresponds to the optimization of the
Lagrangian function $\mathcal{C}_p(w;\lambda)$ using the Adam optimizer. We also denote ex post regrets of buyers and sellers as $\widetilde{\textbf{rgt}}_i^{\text{(B)}}(w),\, \widetilde{\textbf{rgt}}_j^{\text{(S)}}(w)$ and budget-balanced constraint violation as $\widetilde{\textbf{bbp}}(w)$. Then, the gradient of $\mathcal{C}_p$ for fixed $\lambda$ is calculated by

\begin{align*}
    \nabla_w \, C_p(w;\lambda^t) =& -\frac{1}{L} \sum_{\ell=1}^L \Bigl( \sum_i^n (v_{\ell, i}^{\text{(B)}}\sum_j^m \nabla_w \, g_{ij}^w) - \sum_j^m(v_{\ell, j}^{\text{(S)}}\sum_i^n \nabla_w \, g_{ij}^w) \Bigr) \\
                   &+ \sum_{i\in N} \sum_{\ell=1}^L {\lambda_i^\text{(B)}}^{t} q_{\ell, i}^{\text{(B)}} + \rho \sum_{i \in N} \sum_{\ell=1}^L \widetilde{\textbf{rgt}}_i^{\text{(B)}}(w) \, q_{\ell, i}^{\text{(B)}} \\
                   &+ \sum_{j\in M} \sum_{\ell=1}^L {\lambda_j^\text{(S)}}^{t} q_{\ell, j}^{\text{(S)}} + \rho \sum_{j \in N} \sum_{\ell=1}^L \widetilde{\textbf{rgt}}_j^{\text{(S)}}(w) \, q_{\ell, j}^{\text{(S)}} \\
                   &+ {\lambda^\text{(A)}}^{t}\, \nabla_w \, \widetilde{\textbf{bbp}}(w) + \rho \; \widetilde{\textbf{bbp}}\,(w), 
\end{align*}
where

\begin{align*}
    &q_{\ell, i}^{\text{(B)}} = \nabla_w \, \lbrack  \max_{\quad {v^{\prime\text{(B)}}_{\ell, i}}} u_{i}^{\text{(B)}}(v_{\ell, i}^{\text{(B)}}, ({v^{\prime\text{(B)}}_{\ell, i}}, {v^{\text{(B)}}_{\ell, -i}}), v_{\ell}^{\text{(S)}}) - u_{i}^{\text{(B)}}(v_{\ell, i}^{\text{(B)}}, v_{\ell}^{\text{(B)}}, v_{\ell}^{\text{(S)}})  \rbrack, \\
    &q_{\ell, j}^{\text{(S)}} = \nabla_w \, \lbrack \max_{\quad {v^{\prime\text{(S)}}_{\ell, j}}} u_{j}^{\text{(S)}}(v_{\ell, j}^{\text{(S)}}, v_{\ell}^{\text{(B)}}, ({v^{\prime\text{(S)}}_{\ell, j}}, {v^{\text{(S)}}_{\ell, -j}})) - u_{j}^{\text{(S)}}(v_{\ell, j}^{\text{(S)}}, v_{\ell}^{\text{(B)}}, v_{\ell}^{\text{(S)}}) \rbrack.
\end{align*}
The computation of $q_{\ell, i}^{\text{(B)}}, q_{\ell, j}^{\text{(S)}}, \widetilde{\text{rgt}}_i^{\text{(B)}}(w) $, and $ \widetilde{\text{rgt}}_j^{\text{(S)}}(w)$ requires finding an optimal misreport that maximizes the utility of each player. We solve this maximization over misreports using stochastic gradient descent. 
The optimal misreports ${v'_i}^{\text{(B)}}$ (${v'_j}^{\text{(S)}}$) for buyer $i$ (seller $j$) are computed by $R$ gradient updates.

\begin{figure}[t]
\begin{algorithm}[H]
  \caption{DoubleRegretNet Training}
  \label{alg1}
  \begin{algorithmic}[]    
  \renewcommand{\algorithmicrequire}{\textbf{Input:}}
  \REQUIRE Minibatches $\mathcal{S}_1, ..., \mathcal{S}_T$ of size $L$
  \renewcommand{\algorithmicrequire}{\textbf{Parameters:}}
  \REQUIRE $\forall{t}, \rho_t > 0, \gamma > 0, \eta > 0, R \in \mathbb{N}, Q \in \mathbb{N}$
  \renewcommand{\algorithmicensure}{\textbf{Initialize:}}
  \ENSURE $w^0 \in \mathbb{R}^d, \lambda^0 \in \mathbb{R}^n$ 
  \renewcommand{\algorithmicensure}{\textbf{Procedure:}}
  \ENSURE
  \FOR {$t = 0$ \textbf{to} $T$}
  \STATE Receive minibatch $\mathcal{S}_t = \{(v_1^{\text{(B)}}, v_1^{\text{(S)}}), ..., (v_L^{\text{(B)}}, v_L^{\text{(S)}})\}$
  \STATE Initialize misreports ${v'}_{\ell,i}^{\text{(B)}}$ and ${v'}_{\ell,j}^{\text{(S)}}, \forall \ell \in [L], i \in N, j \in M$
    \FOR {$r = 0$ \textbf{to} $R$}
    \STATE $\forall \ell \in [L], i \in N :$
    \quad ${v'}_{\ell, i}^{\text{(B)}} \leftarrow {v'}_{\ell, i}^{\text{(B)}} + \gamma \nabla_{v'_i} \: u_i^{\text{(B)}}(v_{\ell, i}^{\text{(B)}}; ({v'}_{\ell, i}^{\text{(B)}}, {v^{\text{(B)}}_{\ell, -i}}), v_{\ell}^{\text{(S)}})$
    \STATE $\forall \ell \in [L], j \in M :$
    \quad ${v'}_{\ell, j}^{\text{(S)}} \leftarrow {v'}_{\ell, j}^{\text{(S)}} + \gamma \nabla_{v'_j} \: u_j^{\text{(S)}}(v_{\ell, j}^{\text{(S)}}; v_{\ell}^{\text{(B)}}, ({v'}_{\ell, j}^{\text{(S)}}, {v^{\text{(S)}}_{\ell, -j}}))$
    \ENDFOR
    \STATE Compute regret gradient: $\forall \ell \in [L], i \in N, j \in M :$
    \STATE \qquad ${q}_{\ell, i}^{\text{(B)}} = \nabla_w[u_{i}^{\text{(B)}}(v_{\ell, i}^{\text{(B)}}, ({v^{\prime\text{(B)}}_{\ell, i}}, {v^{\text{(B)}}_{\ell, -i}}), v_{\ell}^{\text{(S)}}) - u_{i}^{\text{(B)}}(v_{\ell, i}^{\text{(B)}}, v_{\ell}^{\text{(B)}}, v_{\ell}^{\text{(S)}})]\left.\right|_{w = w^t}$
    \STATE \qquad ${q}_{\ell, j}^{\text{(S)}} = \nabla_w[u_{j}^{\text{(S)}}(v_{\ell, j}^{\text{(S)}}, v_{\ell}^{\text{(B)}}, ({v^{\prime\text{(S)}}_{\ell, j}}, {v^{\text{(S)}}_{\ell, -j}})) - u_{j}^{\text{(S)}}(v_{\ell, j}^{\text{(S)}}, v_{\ell}^{\text{(B)}}, v_{\ell}^{\text{(S)}})]\left.\right|_{w = w^t}$
    \STATE Compute Lagrangian gradient and update $w^t$: 
    \STATE \qquad $w^{t+1} \leftarrow w^t - \eta \nabla_w \, \mathcal{C}_{\rho_t}(w^t, \lambda^t)$
    \STATE Update Lagrange multipliers once in $Q$ iterations: 
      \STATE \qquad \textbf{if} $t$ is a multiple of $Q$
      \STATE \qquad \quad ${\lambda_i^{\text{(B)}}}^{t+1} \leftarrow {\lambda_i^{\text{(B)}}}^{t} + \rho_t \, \widehat{\textbf{rgt}}_i(w^{t+1}), \forall i \in N$ 
      \STATE \qquad \quad ${\lambda_j^{\text{(S)}}}^{t+1} \leftarrow {\lambda_j^{\text{(S)}}}^{t} + \rho_t \, \widehat{\textbf{rgt}}_j(w^{t+1}), \forall j \in M$ 
      \STATE \qquad \quad ${\lambda^{\text{(A)}}}^{t+1} \leftarrow {\lambda^{\text{(A)}}}^{t} + \rho_t \, \widehat{\textbf{bbp}}(w^{t+1})$
      \STATE \qquad \textbf{else} 
      \STATE \qquad \quad $\lambda^{t+1} \leftarrow \lambda^t$ 
  \ENDFOR
  \end{algorithmic}
\end{algorithm}
\end{figure}

\section{Experiments}
\subsection{Settings}
We implemented our framework using the PyTorch deep learning library. 
\footnote{All code is available through the repository at https://anonymous.4open.science/r/deep-opt-double-auction-1C1D/}
We used the tanh activation function at the hidden nodes and Glorot uniform initialization based on existing research.

We used a sample size of $640,000$ valuation profiles for training. 
Player valuations were independently drawn from the uniform distribution $U[0, 1]$.
We ran the augmented Lagrangian solver for a maximum of 80 epochs with a minibatch size of 128. The value of the fixed parameter $\rho$ was initially set to $1.0$ and incremented every two epochs.

The model parameter $w^t$ was updated by the Adam optimizer with a learning rate of $0.001$ for each minibatch, and we performed $R=25$ each player misreport update steps with a learning rate of $0.1$.
We cached the optimal misreport from the current minibatch after 25 updates and used it as the initial value of the misreport for the same minibatch in the next epoch.
We set $Q$ to $100$, so Lagrangian multipliers were updated once every $100$ minibatches.

\subsection{Evaluations}
We used three evaluation indices to measure how much the learnt mechanism satisfies Pareto Efficiency, balanced budget, and incentive compatibility:
\begin{itemize}
    \item social surplus: $\textbf{welfare} = \sum_{i=1}^n (v_i^{\text{(B)}}\sum_{j=1}^m g_{ij}^w - \sum_{j=1}^m(v_j^{\text{(S)}}\sum_{i=1}^n g_{ij}^w)  $
    \item ex post regret: 
    $\textbf{rgt} = \frac{1}{n+m} (\sum_{i=1}^n \widehat{\textbf{rgt}}_i^{(B)} (g^w, p^w)) + \sum_{j=1}^m \widehat{\textbf{rgt}}_j^{(S)} (g^w, p^w))$ 
    \item budget-balanced penalty:
    $\textbf{bbp} = \max \Bigl\{0, \sum_{j=1}^m r_{j}^w(v_\ell^{\text{(B)}}, v_\ell^{\text{(S)}}) - \sum_{i=1}^n p_{i}^w(v_\ell^{\text{(B)}}, v_\ell^{\text{(S)}}) \Bigr\}$.
\end{itemize}
As we introduce randomized trade matching in this study, in addition to these evaluations, we also evaluated the normalized entropy 
to measure the ambiguity of the learnt mechanism:
\begin{itemize}
    \item mechanism ambiguity: 
    $\textbf{entropy} =  - \frac{1}{2n}\sum_{i=1}^n \sum_{j=0}^m \frac{g_{ij} \, \text{log}_2 \, g_{ij}}{\text{log}_2 \, m} - \frac{1}{2m}\sum_{j=1}^m \sum_{i=0}^n \frac{g_{ij} \, \text{log}_2 \, g_{ij}}{\text{log}_2 \, n} $.
\end{itemize}
The closer this normalized entropy is to 0, the more deterministic the trade matching is. The normalized entropy for the existing protocol is surely $0$ due to $g_{ij} \in \{0, 1\}$

\subsection{Results}

\begin{figure*}[t]
  \begin{center}
  \begin{minipage}[b]{0.32\linewidth}
    \centering
    \includegraphics[keepaspectratio, scale=0.5]{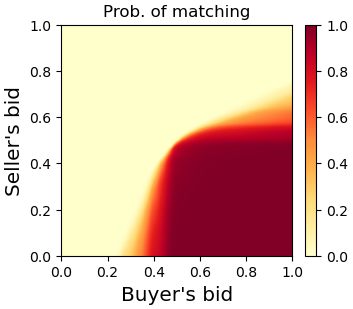}
    {Matching Network}
  \end{minipage}
  \begin{minipage}[b]{0.32\linewidth}
    \centering
    \includegraphics[keepaspectratio, scale=0.5]{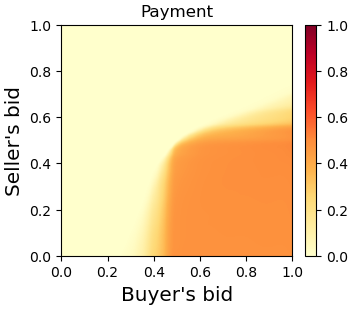}
    {Payment Network}
  \end{minipage}
  \begin{minipage}[b]{0.32\linewidth}
    \centering
    \includegraphics[keepaspectratio, scale=0.5]{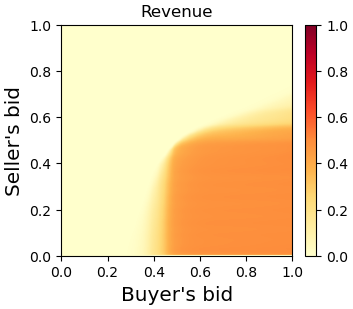}
    {Revenue Network}
  \end{minipage}
  \caption{Output of each network for one buyer and one seller.}
  \label{figure:test_result_1}
  \end{center}
\end{figure*}

\begin{figure*}[t]
  \begin{center}
  \begin{minipage}[b]{0.32\linewidth}
    \centering
    \includegraphics[keepaspectratio, scale=0.5]{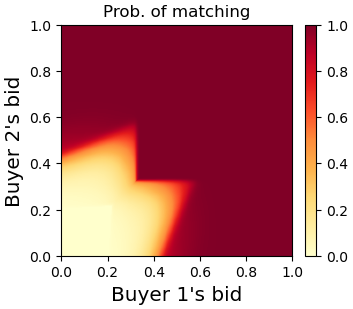}
    {seller bid = $0.2$}
  \end{minipage}
  \begin{minipage}[b]{0.32\linewidth}
    \centering
    \includegraphics[keepaspectratio, scale=0.5]{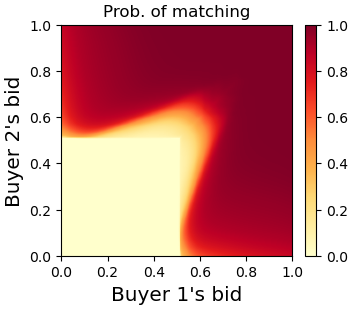}
    {seller bid = $0.5$}
  \end{minipage}
  \begin{minipage}[b]{0.32\linewidth}
    \centering
    \includegraphics[keepaspectratio, scale=0.5]{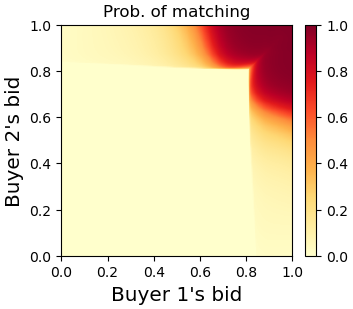}
    {seller bid = $0.8$}
  \end{minipage}
  \caption{Output of matching network for two buyers and one seller.}
  \label{figure:test_result_2}
  \end{center}
\end{figure*}
\begin{figure*}[h]
  \begin{center}
  \begin{minipage}[b]{0.32\linewidth}
    \centering
    \includegraphics[keepaspectratio, scale=0.5]{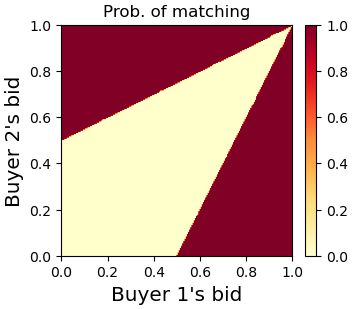}
    {seller bid = $0.2$}
  \end{minipage}
  \begin{minipage}[b]{0.32\linewidth}
    \centering
    \includegraphics[keepaspectratio, scale=0.5]{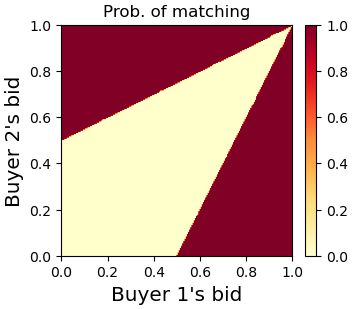}
    {seller bid = $0.5$}
  \end{minipage}
  \begin{minipage}[b]{0.32\linewidth}
    \centering
    \includegraphics[keepaspectratio, scale=0.5]{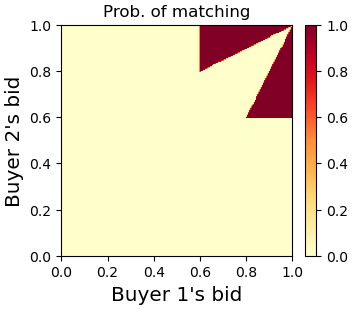}
    {seller bid = $0.8$}
  \end{minipage}
  \caption{Trade matching output of MD protocol for two buyers and one seller.}
  \label{figure:test_result_md}
  \end{center}
\end{figure*}

\begin{figure*}[t]
  \begin{center}
  \begin{minipage}[b]{0.32\linewidth}
    \centering
    \includegraphics[keepaspectratio, scale=0.5]{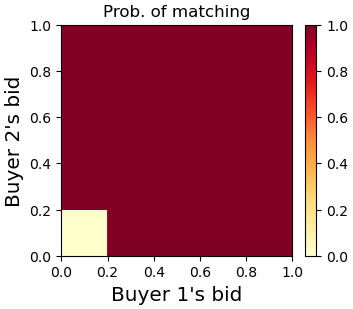}
    {seller bid = $0.2$}
  \end{minipage}
  \begin{minipage}[b]{0.32\linewidth}
    \centering
    \includegraphics[keepaspectratio, scale=0.5]{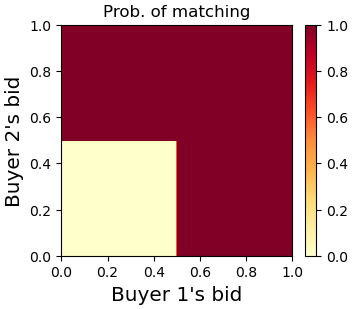}
    {seller bid = $0.5$}
  \end{minipage}
  \begin{minipage}[b]{0.32\linewidth}
    \centering
    \includegraphics[keepaspectratio, scale=0.5]{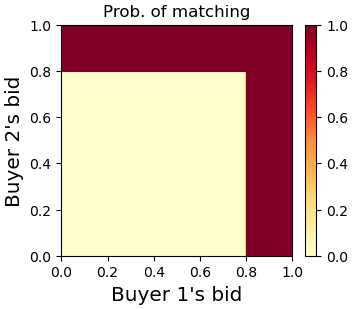}
    {seller bid = $0.8$}
  \end{minipage}
  \caption{Trade matching output of VCG protocol for two buyers and one seller.}
  \label{figure:test_result_vcg}
  \end{center}
\end{figure*}

In our experiment, we assumed a small scale double auction in consideration of the performance evaluation time. 
We first conducted the experiment in the simplest setting with one buyer and one seller.
The test data is an exhaustive combination of values from $0.0$ to $1.0$ in $0.005$ increments.
Figure \ref{figure:test_result_1} shows the outputs of three learnt networks, taking the buyer bid on the horizontal axis and seller bid on the vertical axis, and displaying the corresponding outputs in a color map.
The learnt mechanism indicates that 
a buyer and seller should make a deal if their bids are in the red range of the matching network output, with the price the payment network or the revenue network outputs represented as the orange color (about $0.5$).
We can visually confirm how deterministic the trade matching is.
We also see the learnt mechanism by \textit{DoubleRegretNet} is incentive compatible to some extent considering the Pareto efficient mechanism, where a buyer and seller trade if the buyer bid is greater than or equal to seller bid.

Then, we conducted a similar experiment in the setting with two buyers and one seller.
We applied the exhaustive combinations in the previous setting to valuations of two buyers and fixed the seller valuation to $0.2, 0.5$, or $0.8$.
Figure \ref{figure:test_result_2} shows each network output by color mapping. 
In this setting, one buyer valuation is on the horizontal axis, and the other buyer valuation is on the vertical axis. 
We gained the result that there are no trades if both bids of two buyers are less than the fixed seller bid, which it not counter-intuitive.
To compare the result with existing protocols, Figures \ref{figure:test_result_md} and \ref{figure:test_result_vcg} show double auction results of the MD and VCG protocols for the same setting.
From the viewpoint of Pareto efficiency, the learnt mechanism achieves intermediate social surplus between the MD and VCG protocols in all cases.
The smaller range of transactions than VCG protocol results is due to the consideration of incentive compatibility in \textit{DoubleRegretNet}.
These results suggest the possibility of designing a new mechanism for double auction.

\begin{table*}[t]
    \caption{Results by the proposed \textit{DoubleRegretNet} compared with those by existing protocols.}
    \label{table:test_result}
    \centering
    \begin{tabular}{|c|c|cc|cccc|}
    \hline
        &MD Protocol&  \multicolumn{2}{c|}{VCG Protocol} & \multicolumn{4}{c|}{DoubleRegretNet} \\
    Setting &$\text{welfare}$&$\text{welfare}$&$\text{bbp}$&$\text{welfare}$&$\text{bbp}$&$\text{rgt}$&$\text{entropy}$ \\
    \hline \hline
    $2 \times 2$  & $0.376$ & $0.437$ & $0.220$ & $0.413$ & $0.00002$ & $0.005$ & $0.072$\\
    \hline
    $3 \times 3$  & $0.640$ & $0.703$ & $0.236$ & $0.683$ & $0.00007$ & $0.005$ & $0.027$ \\
    \hline

    $5 \times 5$ & $1.074$ & $1.135$ & $0.227$ & $1.123$ & $0.00015$ & $0.010$ & $0.012$\\
    \hline
    \end{tabular}
\end{table*}

We next measured the performance of the learnt mechanism using four evaluation indices to numerically compare with the MD and VCG protocols.
We tested three settings: two buyers and two sellers ($2 \times 2$), three buyers and three sellers ($3 \times 3$), and five buyers and five sellers ($5 \times 5$).
In the $2 \times 2$ and $3 \times 3$ settings, we used the exhaustive test data, a combination of values from $0.0$ to $1.0$ in $0.1$ increments, that is, $11^4$ valuation profiles for the $2 \times 2$ setting and $11^6$ valuation profiles for the $3 \times 3$ setting.
In the $5 \times 5$ setting, we instead conducted the experiment with $10,000$ sampled test datasets.
Table \ref{table:test_result} shows the expected values of evaluations in each setting.
We observed a similar tendency in the results of all settings.
The social surplus \textbf{welfare} of the learnt mechanism does not exceed that of the VCG protocol, which guarantees Pareto efficiency; however, it exceeds that of the MD protocol.
The budget-balanced penalty(\textbf{bbp}) of the learnt mechanism is nearly zero, which means that the mechanism almost satisfies balanced budget, while the \textbf{bbp} of the VCG protocol is so large that it could bring a deficit to the auctioneer.
Also, ex post regret \textbf{rgt} of the mechanism obtained from \textit{DoubleRegretNet} is very small, which indicates that the mechanism does not give buyers and sellers any incentive to deliberately misreport.
In addition, the normalized entropy is small enough to confirm that we gain the deterministic mechanism to some extent.
Overall, \textit{DoubleRegretNet} successfully learns an individual rational mechanism that is more economically efficient than the MD protocol with balanced budget and incentive compatibility mostly satisfied.

\section{Conclusion}
This paper presented a neural network architecture \textit{DoubleRegretNet} to learn the mechanism for double auction through deep learning. 
This consists of three components: a matching network to determine the pair of making a deal, a payment network to compute how much each buyer should pay, and a revenue network to compute how much each seller should receive.
We then constructed \textit{DoubleRegretNet} to guarantee individual rationality and output equivariant results to any permutation of inputs.
We conducted experiments to confirm the learnt mechanism by \textit{DoubleRegretNet} after formulating double auction as an optimization problem.
Consequently, we obtained a mechanism that brings more utilities to players than the MD protocol and almost satisfies balanced budget and incentive compatibility besides individual rationality.
Our future work will aim to extract the concrete interpretability from the learnt mechanism and to extend to more complicated auction settings.

\nocite{*}
\bibliography{report}

\end{document}